\documentclass{aa}
\usepackage{epsfig}

\def\ltsima{$\; \buildrel < \over \sim \;$}
\def\simlt{\lower.5ex\hbox{\ltsima}}
\def\gtsima{$\; \buildrel > \over \sim \;$}
\def\simgt{\lower.5ex\hbox{\gtsima}}

\begin{document}
   \thesaurus{03 (13.25.2;11.14.1;12.04.2)}
  \title{Testing current synthesis models of the X--ray background}

   \author{R. Gilli M. 
                \inst{1,2}
   \and    M. Salvati
		\inst{3}		
   \and    G. Hasinger
                \inst{2}
}
   \offprints{R. Gilli}

  \institute {Dipartimento di Astronomia e Scienza dello Spazio, 
Universit\`a di Firenze, Largo E. Fermi 5, I--50125 Firenze, Italy 
(gilli@arcetri.astro.it)
     \and Astrophysikalisches Intitut Potsdam, An der Sternwarte 16, 
D--14482 Potsdam, Germany (ghasinger@aip.de)    
     \and   Osservatorio Astrofisico di Arcetri, Largo E. Fermi 5,
I--50125 Firenze, Italy (salvati@arcetri.astro.it)}

   \date{Received / Accepted }

\titlerunning{Testing current synthesis models of the XRB}
\authorrunning{R. Gilli et al.}

   \maketitle

   \begin{abstract}

We present synthesis models of the X--ray background where 
the available X--ray observational constraints are used to 
derive information on the AGN population properties. 
We show the need for luminous X--ray absorbed AGNs, the QSO2s,
in reproducing the 2--10 keV source counts at relatively bright fluxes.
We compare a model where the evolution of absorbed AGNs 
is faster than that of unabsorbed ones, with a standard model where 
absorbed and unabsorbed AGNs evolve at the same rate. It is found 
that an increase by a factor of $\sim 2$ from $z=0$ to $z\sim1.3$ 
in the ratio between absorbed and unabsorbed AGNs would
provide a significant improvement in the data description.
Finally, we make predictions on the AGNs to be observed in deep X--ray
surveys which contain information on the AGN space density at high redshift. 

   \end{abstract}


%

\section{Introduction}

Recent surveys have provided strong evidence that the extragalactic 
X--ray background (XRB) is mainly due to the integrated emission
of single sources. In the soft 0.5--2 keV energy range about 70--80\% of the 
XRB intensity has been resolved by ROSAT at a limiting flux of
$10^{-15}$ erg cm$^{-2}$ s$^{-1}$ (Hasinger et al. \cite{hasinger}). 
The resolved
fraction increases up to $\sim 90\%$ when adding the Chandra results
down to a flux of $2.3\times10^{-16}$ erg cm$^{-2}$ s$^{-1}$ (Mushotzky 
et al. \cite{musho} ; Giacconi et al. \cite{giacconi}). 
In the hard 2--10 keV energy range
ASCA and BeppoSAX resolved about 25--30\% of the XRB (Ueda et al. \cite{ueda}; 
Della Ceca et al. \cite{rdc2}; Giommi, Perri \& Fiore \cite{giommi}). 
When adding Chandra data (Mushotzky et al. \cite{musho}; 
Giacconi et al. \cite{giacconi}) more than 60\% of the hard XRB is resolved.
Most of the ROSAT and ASCA sources have been optically identified as 
Active Galactic Nuclei (AGNs; Schmidt et al. \cite{schmidt2}; 
Akiyama et al. \cite{ak00}, hereafter Ak00). 
The optical identifications for the Chandra sources are not complete, however 
AGNs seem to constitute the main population (Barger et al. \cite{barger}). 
These results have confirmed the predictions of AGN synthesis
models of the XRB, originally proposed by Setti \& Woltjer (\cite{setti}) and 
subsequently refined into more detailed variants 
(e.g. Madau, Ghisellini \& Fabian \cite{madau}; 
Comastri et al. \cite{comastri}); 
the XRB spectrum is reproduced by summing the 
contribution of unobscured and obscured AGNs.
In these models, the hard spectrum of obscured AGNs is crucial in reproducing 
the XRB spectral shape above 2 keV.

The existence of a population of luminous absorbed AGNs, the QSO2s, is still 
a debated issue. Some QSO2s could be hidden in the Ultraluminous IRAS 
Galaxies (ULIRGs; Kim \& Sanders \cite{kim}). Indeed, many ULIRGs do show AGN 
signatures in their spectra (Veilleux, Kim \& Sanders \cite{veilleux}).
Also, examples of luminous absorbed AGNs could be represented by 
ROSAT sources with very red colours ($R-K\ga5$)  detected at faint
fluxes in the Lockman Hole (Hasinger et al. \cite{hasinger2}). Finally,
another interesting possibility, which however is still to be verified, 
is that QSO2s appear as normal blue, broad lined QSOs in the optical.   
While luminous AGNs with only narrow optical lines are not commonly
observed (see e.g. Halpern, Turner \& George \cite{halpern}), ASCA and 
BeppoSAX have found luminous AGNs with a hard X--ray spectrum showing broad 
lines in the optical (Ak00; Fiore et al. \cite{fiore}).
If the X--ray hardness is due to cold gas absorption, one could 
therefore suspect that these objects have different dust--to--gas
properties with respect to low luminosity AGNs, where X--ray absorption
is generally coupled with an optical type 2 (only narrow lines) 
classification (Maiolino et al. \cite{maio3}; 
Risaliti et al. \cite{risaliti2}).

Recent works (Gilli et al. \cite{gilli}, hereafter Paper I; 
Pompilio et al. \cite{pompilio}) 
suggested that the evolution of obscured AGNs might be different from that 
of unobscured ones, with the number ratio R between obscured and unobscured 
AGNs increasing with redshift. Observational data seem to confirm this
suggestion. Indeed, Reeves \& Turner (\cite{reeves}) found that in a sample 
of radio 
loud and radio quiet QSOs observed by ASCA the fraction of X--ray 
absorbed AGNs increases with redshift.

The density of high redshift AGNs has been determined in the optical 
(Schmidt, Schneider \& Gunn \cite{schmidt}) and in the radio 
(Shaver et al. \cite{shaver}) 
bands. It is found that the AGN density declines above $z\sim3$, as
recently confirmed by results from the Sloan Digital Sky Survey 
(Fan et al. \cite{fan}). In the X--rays the situation is uncertain. 
The behaviour of the soft X--ray Luminosity Function 
(XLF) derived by Miyaji, Hasinger \& Schmidt (\cite{miyaji}, hereafter Mi00a) 
from ROSAT data is indeed consistent with a constant AGN density above 
$z\sim2$. 

In this work we examine XRB synthesis models under 
different assumptions in the input parameters. We check the differences in 
reproducing all the 
available X--ray constraints between a standard model where the ratio R 
does not evolve with the redshift and a model where absorbed AGNs evolve 
more rapidly than unabsorbed ones. We analyze the effects of 
neglecting QSO2s when the XRB intensity is reproduced only with unobscured 
AGNs and low luminosity absorbed AGNs. Finally, we compare two alternative 
scenarios where the density of AGNs at high redshift is constant or declines 
above $z\sim3$. 

Throughout this paper the deceleration parameter and the Hubble constant 
are given the values $q_{0}=0.5$ and $H_{0}=50$ km s$^{-1}$ Mpc$^{-1}$.

\section{The models}

\subsection{The spectra}

The input AGN spectra have been divided into unabsorbed 
AGNs and absorbed AGNs with log$N_{\rm H}$=21.5,22.5,23.5,24.5,25.5, 
as described in Paper I. With respect to Paper I we introduced a change 
in the soft excess of unabsorbed sources to get a better agreement with 
observational data. 
Indeed, the soft excess is not a general property of the 
spectra of unobscured AGNs. Below $z\sim0.3$, where a soft excess is 
observable in the ASCA band, about 50\% of QSOs (George et al. \cite{george}; 
Reeves \& Turner \cite{reeves}) and 20\% of the Seyfert 1s 
(Reynolds \cite{reynolds})
are found to have a soft excess. In the Reeves \& Turner (\cite{reeves}) 
sample the 
median ratio between the soft excess component and the primary power law in 
the 0.5--2 keV band is 0.19. We have then approximated this situation by 
considering a soft excess in {\it all} the unabsorbed AGNs with a 
ratio lower by a factor of $\sim 2$ than the median value of Reeves \& Turner 
(\cite{reeves}). This soft excess is parametrized with a power law with 
photon index $\Gamma=2.4$ below 1 keV.

Another difference with respect to Paper I is in the spectrum of AGNs 
absorbed with log$N_{\rm H}$=24.5, where we have now introduced an 
approximated correction for Compton scattering.
We interpolated the AGN spectra of Matt, Pompilio \& La Franca (\cite{matt}) 
(see their Fig.~4) deriving a function to correct from the case where only 
photoelectric absorption is considered.
When Compton scattering is added, the intensity of the spectrum of 
AGNs with log$N_{\rm H}$=24.5 decreases by a factor of $\sim 2$ at 30 keV.
We note that the effect of Compton scattering is negligible for the 
other classes of AGNs in our model with lower absorption. Following 
Paper I, we have adopted a pure reflection spectrum for sources with 
log$N_{\rm H}$=25.5 since no transmitted component is expected at this
absorption level.
We remind that a soft component has been considered in the input 
spectra of absorbed AGNs, parametrized with a power law with $\Gamma=2.3$, 
normalized to be 3\% of the primary de--absorbed power law
at 1 keV.

\subsection{The number of obscured AGNs}

In our models we fix the local ratio between absorbed and unabsorbed AGNs
to R=4. As discussed in Paper I this value is derived by Maiolino \& Rieke 
(\cite{maio}) as the local ratio between Seyfert 2 and Seyfert 1 galaxies in 
the optical.
Since in the local Universe most of the Seyfert 1 galaxies are unabsorbed
in the X--rays (Schartel et al. \cite{schartel}), R=4 can be used in the 
X--ray domain as well. 
We also adopt R=4 for high luminosity AGNs, although this number is
uncertain. 
The distribution of the absorbing column densities is assumed to be that 
derived by Risaliti, Maiolino \& Salvati (\cite{risaliti}), 
where $\sim 50\%$ of the objects have log$N_{\rm H}\leq 24$.

We will explore a standard model (model A) where the ratio R=4 does not 
change with redshift and a model (model B) where the ratio R increases with
redshift as follows:
\begin{eqnarray}
\begin{array}{ll}
R(z)=4(1+z)^p& z<z_{cut}\\
R(z)=R(z_{cut})=10& z\geq z_{cut}\;\\
\end{array}
\nonumber
\end{eqnarray}
where $z_{cut}$ is the redshift at which AGNs stop evolving.
The absorption distribution is assumed not to vary with
redshift and/or luminosity. The ratio R is assumed not to vary with
luminosity. 

\subsection{The XLF}

The contribution of unabsorbed and absorbed AGNs to the XRB is
evaluated by integrating the XLF of unabsorbed and absorbed 
objects, respectively. The XLF of absorbed AGNs is completely unknown, 
while the XLF of unabsorbed AGNs is usually assumed to coincide with the 
XLF of broad line AGNs. 
As stated in the previous Section, a mismatch between the optical and the 
X--ray classification is sometimes observed, 
therefore considering the XLF of broad line AGNs as the XLF of X--ray 
unabsorbed AGNs cound not be appropriate.
In our model we refer to the XLF derived by Mi00a
from a sample of $\sim 700$ AGNs detected by ROSAT, 
which includes all those objects that are classified as AGNs, irrespectively
of any distinction into type 1 and type 2 subclasses.
These authors find that a Luminosity Dependent Density Evolution (LDDE), 
where the evolution rate drops with decreasing AGN luminosity, provides a 
good representation of the data, while a Pure Luminosity Evolution (PLE) or a 
Pure Density Evolution (PDE) are rejected. 

Although derived in the soft band, and then mainly populated by X--ray 
unabsorbed AGNs, the XLF of Mi00a is expected to include also 
some absorbed AGNs, which could show up because of 
their soft components or, for sources at high redshift, because of the
K--correction. Also, the Mi00a XLF is given in the observed 
and not in the rest frame 0.5--2 keV band.
We have therefore adopted the following approach to obtain a self--consistent 
model. First we have considered the LDDE1 representation of
Mi00a, where we have arbitrarily modified some parameters
to get a rest frame 0.5--2 keV XLF for unabsorbed AGNs only. 
The rest frame 0.5--2 keV XLFs for the AGNs in each absorption class of 
our model are then obtained by multiplying the XLF of unabsorbed AGNs by 
R$\times f_i$, where $f_i$ is the fraction of AGNs in each absorption 
class according to the distribution of Risaliti et al. (\cite{risaliti}). 
The modified XLF parameters were then tuned 
to obtain a good fit to the XRB by summing the contribution of 
unobscured and obscured AGNs.

Once the XRB spectrum has been fitted, we calculate the {\it observed} XLF 
for each absorption class in our model, applying the corrections for 
the absorption and the K--correction. Finally we sum up all the observed 
XLF for the individual absorption classes to derive the total observed XLF
predicted by the model, which can be compared with the binned XLF data 
quoted by Miyaji, Hasinger \& Schmidt (\cite{miyaji2}).

\section{Results}

\subsection{The XRB spectrum}

The parameters of the LDDE1 description (Mi00a) which we
tune in our models are $\gamma_1$, $L_a$ and $z_{cut}$. They 
represent respectively the slope of the low luminosity part of the XLF, 
the luminosity below which the density evolution rate starts to drop, 
and the redshift at which the evolution stops.
Their values for models A and B are listed in 
Table~1. In model B we fixed $R(0)=4$ and $R(z_{cut})=10$, therefore the two
models have the same number of free parameters.
Following Paper I, the XLF was integrated between 
$L_{0.5-2}=10^{41}-10^{49}$ erg s$^{-1}$ and $z=0-4.6$.
Given our assumptions the XLF of absorbed AGNs in model B evolves
according to $(1+z)^{p_{abs}}$, where $p_{abs}=6.19$.
The fit to the XRB spectrum is shown in Fig.~1.
The parameters were tuned in order to obtain the same XRB intensity at 
5 keV for both models. In the 2--10 keV band the uncertainties in the 
measured level of the XRB are of the order of 30--40\% between different
missions (see Vecchi et al. \cite{vecchi}). We chose to fit the XRB intensity 
of the ASCA data (Miyaji et al. in preparation; 
Gendreau et al. \cite{gendreau}) which represent a
median level among the lowest and highest intensities measured by
HEAO--1 (Gruber \cite{gruber}; Gruber et al. \cite{gruber2}) and BeppoSAX 
(Vecchi et al. \cite{vecchi}), respectively.
As seen in Fig.~1 both models provide a good fit to 
the XRB spectrum and are identical in the 3--20 keV energy
range. As in Paper I, we have also added the contribution of 
clusters of galaxies, which is non negligible only below $\sim 3$ keV.
Given the higher fraction of absorbed AGNs in model B, 
the resulting XRB spectrum is harder than that calculated according to 
model A. 

\begin{figure}
\epsfig{file=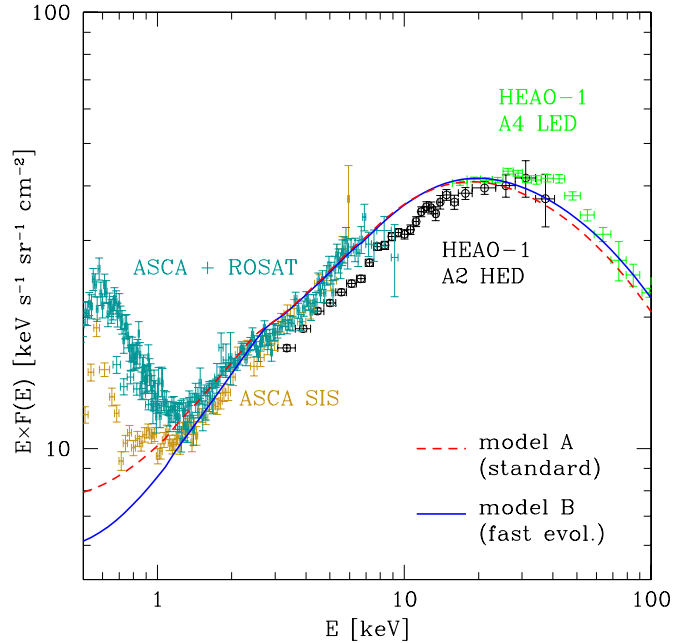,  width=9.cm, height=9.cm,
angle=0}
\caption{The fit to the XRB given by model A and model B. 
The data below $\sim 10$ keV are
from ASCA SIS (Gendreau et al. \cite{gendreau}) and from a combined 
ASCA GIS + ROSAT PSPC analysis in the ASCA Large Sky Survey region 
(Miyaji et al. in prep.). 
The data in the 3--40 keV and 15--100 keV ranges are respectively from the
A2 HED and A4 LED detectors on board HEAO--1 
(Gruber \cite{gruber}; Gruber et al. \cite{gruber2}).}
\end{figure}

\begin{table}
\caption{Model parameters compared with those of Mi00a.}
\label{models}
\begin{tabular}{llll|l}
\hline \hline   
Parameter& model A& model B& model C&Mi00a\\
&QSO2s&QSO2s&no QSO2s& \\
\hline 
$\gamma_1$& 0.62& 0.62& 0.62& $0.75^{+0.15}_{+0.15}$\\
log$L_a$& 44.3& 44.65& 44.4& 44.1$^a$\\
$z_{cut}$& 1.54& 1.32& 1.38& $1.57^{+0.15}_{+0.15}$\\
\hline
R(0)& 4& 4& 4$^b$& \\
R($z_{cut}$)& 4& 10& 10$^b$& \cr\hline
\end{tabular}

Errors in the Mi00a values are at 90\% confidence level.\\ 
$^a$Fixed value by Mi00a.\\
$^b$Ratio at low luminosity. At high luminosity R(z)=0.
\end{table}

\begin{figure*}
\hspace{-1.2cm}
\epsfig{file=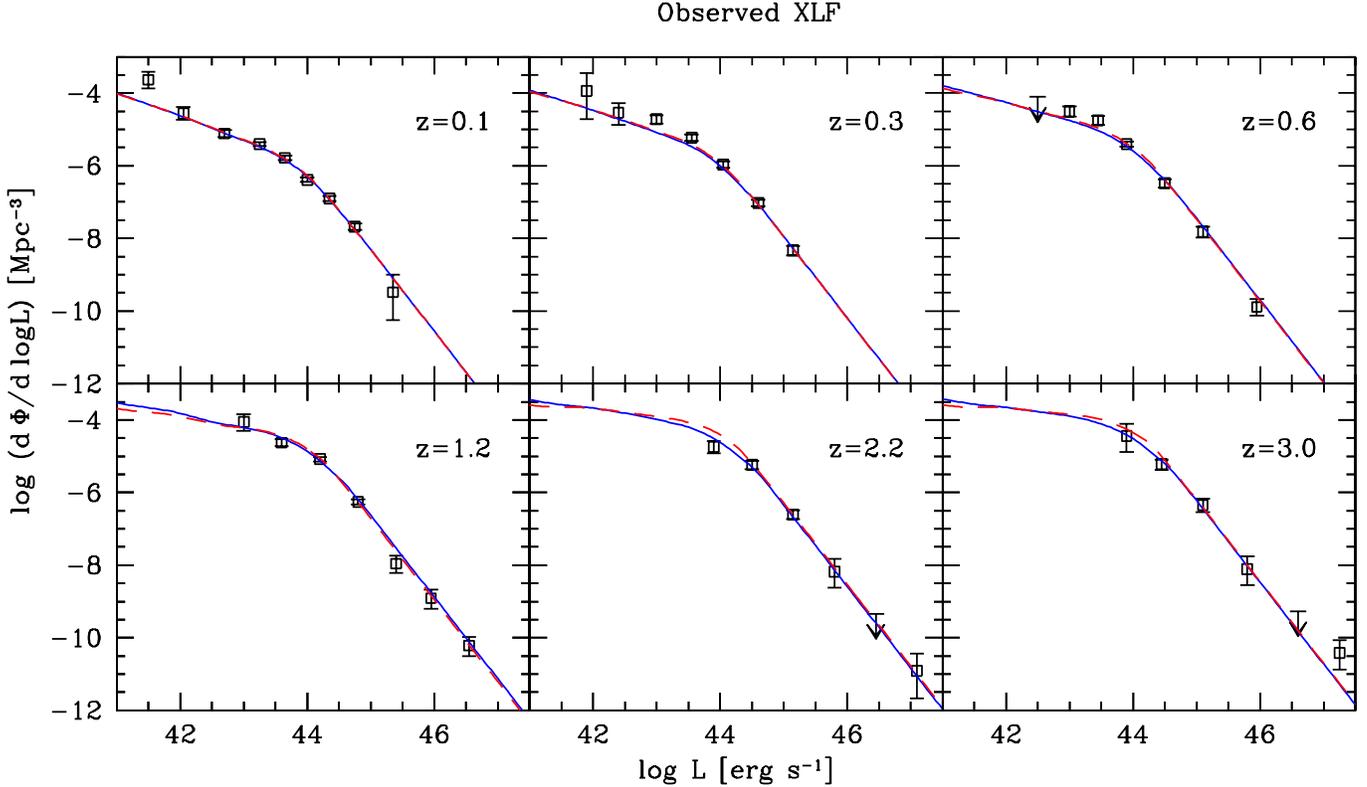,  width=20.cm, height=11.cm,
angle=0}
\caption{Comparison between the total observed AGN XLF 
(Miyaji et al. \cite{miyaji2}) 
and those predicted by model A (dashed) and model B (solid line) by 
summing up the contribution of unabsorbed and absorbed AGNs.} 
\end{figure*}

\subsection{The soft XLF}

Starting from the assumed 0.5--2 keV rest frame XLF and after applying 
corrections for the absorption 
and the redshift we have reconstructed the {\it total} AGN XLF in the 
{\it observed} 0.5--2 keV band according to the two models, which can 
be compared with the data of Miyaji et al. (\cite{miyaji2}). 
In Fig.~2 we show this comparison. 
Both models provide a good fit to the data. The predictions are 
almost identical at low redshift, with some differences in the two 
highest redshift bins.
We have performed a $\chi^2$ test to verify the goodness of the
fit; the results for the two models are shown in Table~2.

\subsection{The X--ray source counts}

We have then compared the predictions of the competing models with the 
source counts observed in the 0.5--2 keV and 2--10 keV energy bands.
In our calculation we have considered the fact that 
the observed logN--logS are derived from samples of objects 
limited in count--rate rather than in flux.
Since X--ray instruments have higher sensitivity at low  
energies, soft sources can be detected at fainter fluxes with respect 
to hard sources. When converting from count rates to fluxes a single  
conversion factor is usually assumed in published works
(corresponding to an average spectrum).
This implies that the flux of a source harder or softer than the average 
one is underestimated or overestimated, respectively.
Therefore, since the number density of sources increases with decreasing flux, 
at any measured flux in the logN--logS the ratio between absorbed and 
unabsorbed AGNs is underestimated.
To cope with this situation we have therefore introduced into our logN--logS 
calculations the same instrumental bias.
Correction for the ROSAT PSPC and ASCA SIS0 effective areas
have been considered in the 0.5--2 keV and 2--10 keV band, respectively.
The ROSAT PSPC correction has been also adopted when calculating the 
predicted XLF shown in Fig.~2.
The inclusion of the instrumental effects is
discussed in detail in the Appendix. 

\begin{figure}
\epsfig{file=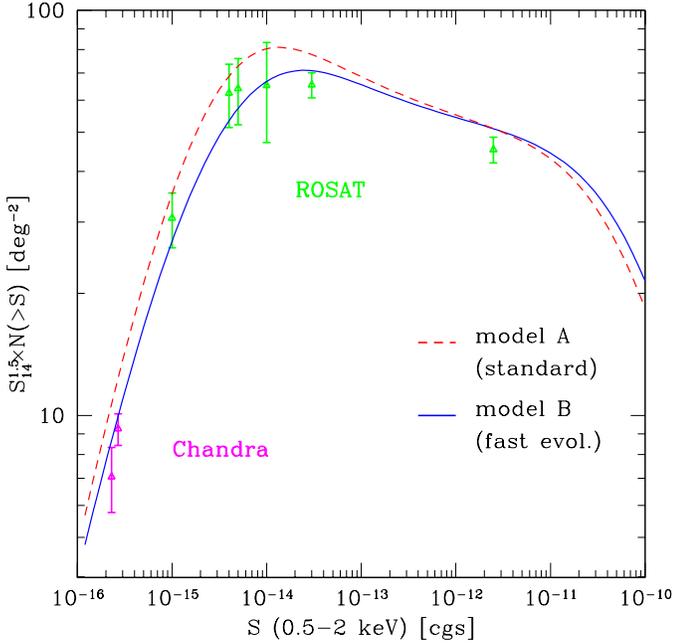,  width=9.cm, height=9.cm, angle=0}
\caption{The predictions of model A and B compared with the
0.5--2 keV source counts. In this and in the
following Figures, source counts are plotted as $S_{14}^{1.5}\times       
N(>S)$, where $S_{14}$ is the flux in units of $10^{-14}$ erg s$^{-1}$
cm$^{-2}$, and errorbars refer to $1\sigma$ uncertainties. 
With decreasing flux, the data are derived from Mi00a,
Mason et al. (\cite{mason}), Bower et al. (\cite{bower}), Mi00a, 
Zamorani et al. (\cite{gz}), 
Hasinger et al. (\cite{hasinger}), Giacconi et al. (\cite{giacconi}), 
and Mushotzky et al. (\cite{musho}). 
The 5 points at higher fluxes contain only AGNs, the remaining three points
at lower fluxes are total counts.}
\end{figure}

\begin{figure}
\label{hard}
\epsfig{file=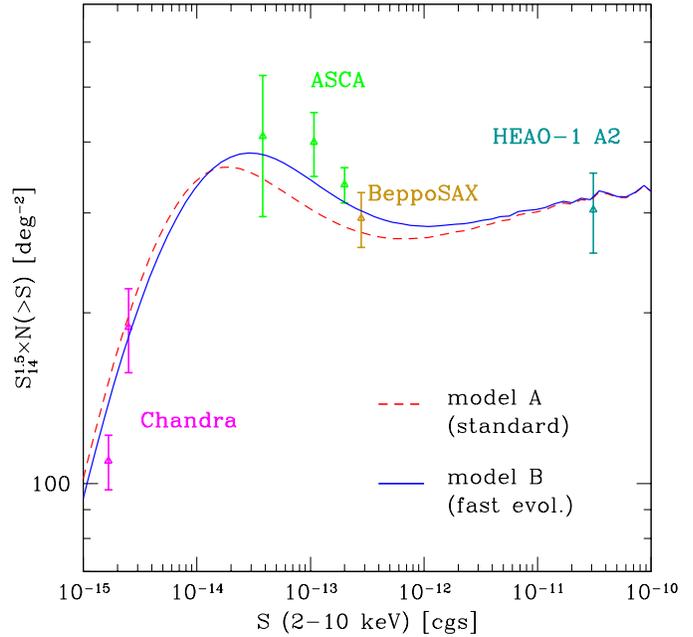, width=9.cm, height=9.cm, angle=0}
\caption{The predictions of model A and B (considering AGNs plus clusters of
galaxies) compared with the total 
2--10 keV counts. 
With decreasing flux, the data are from Piccinotti et al. (\cite{picci}), 
Giommi et al. (\cite{giommi}), Ueda et al. (\cite{ueda}), 
Cagnoni et al. (\cite{cagnoni}), Ogasaka et al. (\cite{ogasaka}), 
Mushotzky et al. (\cite{musho}), 
and Giacconi et al. (\cite{giacconi}).}
\end{figure}

The comparisons of the model predictions with the soft and the hard counts 
are shown in Fig.~3 and 4, respectively.


We performed a $\chi^2$ test on the soft counts to evaluate statistically 
the differences between model A and B. Since the datapoints of Mi00a
are not independent, we have considered literature data
derived from different ROSAT surveys in addition to the Chandra data.
The survey areas adopted to calculate the AGN densities 
have been corrected for incompleteness in the optical identifications.
All the ROSAT AGN densities used in the $\chi^2$ test and shown in Fig.~3
are in agreement within $1\sigma$ with those of Fig.~6 of Mi00a 
apart for the one at $3\times10^{-14}$ erg cm$^{-2}$ s$^{-1}$ derived from
the RIXOS survey (Page et al. \cite{page}; Mason et al. \cite{mason}), 
which is lower by 30\% ($4\sigma$).
Since at the faint Chandra fluxes the optical identifications are scarce,
we compare the models with the total number of X--ray sources. 
The results of the $\chi^2$ test are shown in Table~2.
 


Model B provides a better agreement also with the hard data with respect 
to model A, despite of discrepancies in
the density of hard sources of the order of 20\% between ASCA and 
BeppoSAX data and 40\% between Chandra results from different surveys.
The curves plotted in Fig.~4 include also the contribution of clusters
of galaxies (which is negligible below $\sim 10^{-14}$
erg cm$^{-2}$ s$^{-1}$). At high fluxes the AGN density predicted by 
the model is in agreement with that measured by Piccinotti et al. 
(\cite{picci})\footnote{In Fig.~4 we plot the total source counts
also for the Piccinotti sample, contrary to Paper I where we plotted
only the AGN counts.}

We performed a $\chi^2$ test over a series of total hard 
counts derived from separate surveys (shown in Fig.~4) and then independent 
from each other.
The results are shown in Table~2. 

\begin{table}
\caption{$\chi^2$ test on the XLF and source counts.}
\label{stat}
\begin{tabular}{lll}
\hline \hline   
Data set&  \multicolumn{2}{c}{$\chi^2/dof$}\\
\hline
& model A& model B\\
\hline 
0.5--2 keV XLF& 54.4/38& 49.0/38\\
0.5--2 keV counts& 34.3/8& 8.3/8\\
2--10 keV counts& 20.5/7& 8.5/7\cr
\hline
\end{tabular}
\end{table}

We have verified that the two models are in agreement with the $N_{\rm H}$
distribution observed in the Piccinotti sample 
(Schartel et al. \cite{schartel}).
Since R(0)=4 for both models and since the Piccinotti sample is dominated
by local AGNs, the $N_{\rm H}$ distributions predicted by the two
models are identical.  

\begin{figure}
\hspace{-1.5cm}
\epsfig{file=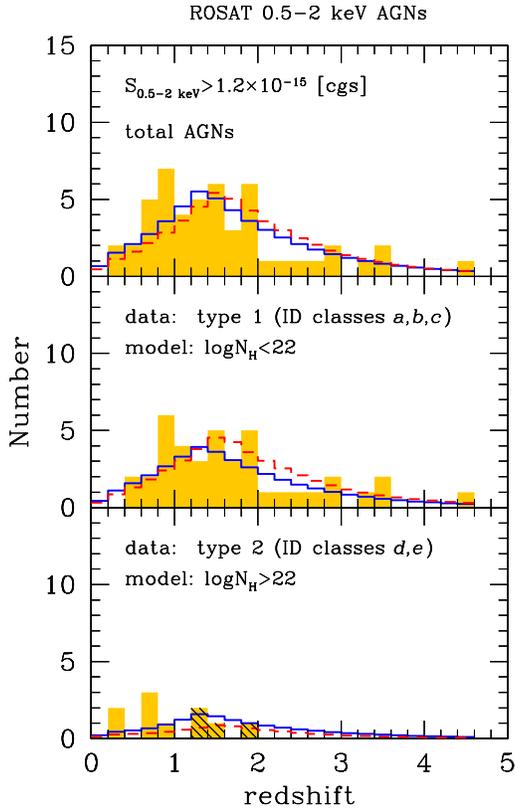,  width=11.cm, height=11.cm, angle=0}
\caption{The redshift distribution predicted by model A (dashed) and B (solid)
compared with the AGN sample from the ROSAT Ultradeep HRI Survey 
(Hasinger et al. \cite{hasinger2}). The shaded histogram refers to AGNs 
having only a photometric redshift estimate.}
\end{figure}

\begin{figure}
\hspace{-1.5cm}
\epsfig{file=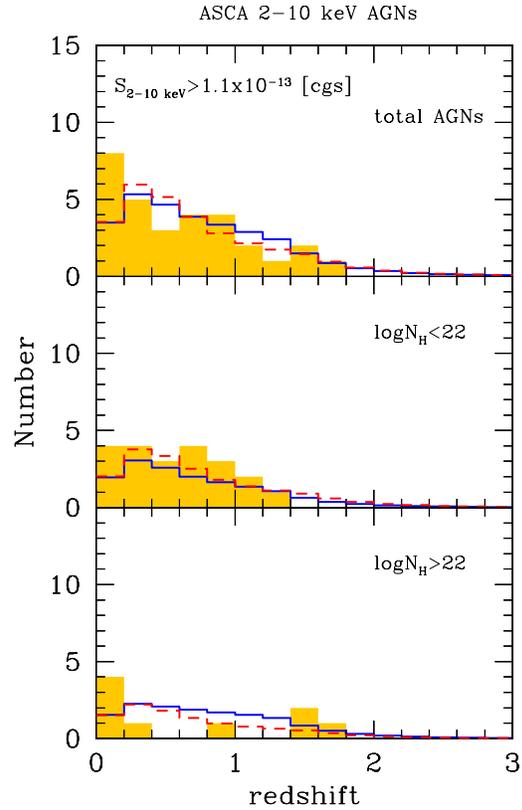,  width=11.cm, height=11.cm, angle=0}
\caption{The redshift distribution predicted by model A (dashed) and B (solid)
compared with the AGN sample from the ASCA Large Sky Survey (Ak00). Four
blue, broad lined QSOs at $z>0.8$ with hard X--ray spectra have been 
included among the AGNs with log$N_{\rm H}>22$.}
\end{figure}

\subsection{Redshift distributions}

The models have been also checked against the redshift 
distribution of AGNs in flux limited samples.
We first consider the data from the ROSAT Ultradeep HRI Survey 
in the Lockman Hole (Hasinger et al. \cite{hasinger2}; 
Lehmann et al. in prep.). 
With an exposure of 1 Msec the achieved
0.5--2 keV limiting flux is $1.2\times10^{-15}$ erg s$^{-1}$ cm$^{-2}$.
We consider only the subsample of sources with an offset angle from
the aimpoint below 12 arcmin, in order to avoid vignetting effects and 
then non uniform limiting fluxes across the field. 
The optical identifications of this sample are almost complete and 50 AGNs
have been identified: 40 of them have 
broad lines (ID classes {\it a,b,c} according to 
the classification scheme of Schmidt et al. \cite{schmidt2}), while 10 have 
no broad 
lines (ID classes {\it d,e}). Four AGNs without broad lines have only a 
photometric redshift estimate, which, if proven to be correct, 
would place them at $z>1$. 

When comparing the model with the data, corrections for the HRI effective 
area have been introduced into the calculations. 
Since the HRI sources have been detected by using the full
HRI band 0.1--2.4 keV, when converting from the count rate 
to a 0.5--2 keV flux, the conversion factor varies by $\pm 40\%$ for 
photon indices $\Gamma=2.0\pm0.7$. This introduces a strong bias 
against hard (absorbed) sources which is introduced in our computations
(see details in the Appendix).
Besides the redshift distribution for total AGNs we calculated the 
distributions for the AGNs with log$N_{\rm H}<22$ and log$N_{\rm H}>22$ separately,
which have been compared with those of AGNs with ID classes {\it a,b,c} and
{\it d,e}, respectively (Fig.~5).
Given the possible mismatch between optical and X--ray classification 
mentioned in 
Section 1, this comparison might not be strictly appropriate. However,
we consider this as a first order check to the model. 
The ratio between AGNs with log$N_{\rm H}<22$ and log$N_{\rm H}>22$ predicted by 
model A and B is 42/8 and 36/14 respectively, to be compared with the ratio
40/10 between AGNs in the {\it a,b,c} classes and AGNs in the {\it d,e} 
classes (the total number of AGNs is normalized to the observed one). 
We have also calculated 
the probability that the observed sample has been drawn from the predicted 
distribution according to a Kolmogorov--Smirnov (KS) test. The calculated
probabilities are quoted in Table~3. Model B is slightly favored by the test.

\begin{table}
\caption{Kolmogorov--Smirnov test on the redshift distributions.}
\label{ks}
\begin{tabular}{lll}
\hline \hline   
Data set&  \multicolumn{2}{c}{$P_{KS}$}\\
\hline
& model A& model B\\
\hline 
ROSAT total& 0.02& 0.23\\
log$N_{\rm H}<22$& 0.26& 0.90\\
log$N_{\rm H}>22$& 0.01& 0.03\\
\hline
ASCA total& 0.43& 0.23\\
log$N_{\rm H}<22$& 0.35& 0.33\\
log$N_{\rm H}>22$& 0.35& 0.24\\
\hline
\end{tabular}
\end{table}
 
Then we compared the predictions of the models with the redshift 
distributions
of AGNs in the 2--10 keV band. We consider the hard sample detected by the 
SIS instrument in the ASCA Large Sky Survey (Ueda et al. \cite{ueda}) 
at a limiting flux of $10^{-13}$ erg cm$^{-2}$ s$^{-1}$ between 2--10 keV, 
which has been completely identified by Ak00.
They find 30 AGNs in a sample of 34 objects (plus 2 clusters of galaxies, 
1 star and 1 unidentified source). These authors also estimate a
column density for the AGNs from a hardness ratio analysis, assuming an
intrinsic spectral index $\Gamma=1.7$. The ratio between AGNs with 
log$N_{\rm H}<22$ and log$N_{\rm H}>22$ is found to be 21/9.
On the basis of their hardness ratios also 4 normal, blue and broad 
lined QSOs with $z>0.8$ are considered to have log$N_{\rm H}>22$.
The unidentified source has a very hard X--ray spectrum ($\Gamma<1$)
and is presumably an obscured AGN.
The predicted ratio N(log$N_{\rm H}<22$)/N(log$N_{\rm H}>22$) 
is 19/11 and 15/15 for model A and B, respectively 
(the total number of AGNs is normalized to the observed one).
The comparison between the observed and predicted redshift distributions
for the Ak00 sample is shown in Fig.~6.
The probability that the sample is drawn from the calculated distributions
is evaluated with a KS test and quoted in Table~3. Here the two models
perform equally well.

%

\subsection{QSO2s}

\begin{figure}
\epsfig{file=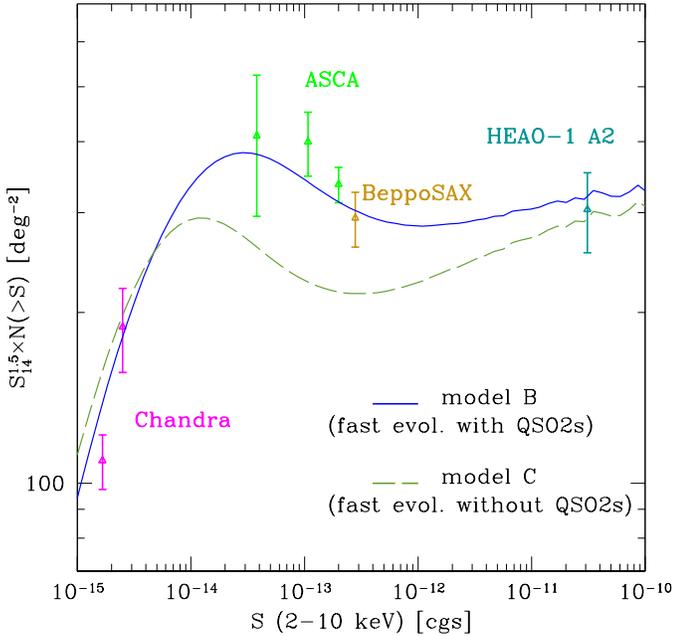, width=9.cm, height=9.cm, angle=0}
\caption{The predictions of model B and C (considering AGNs plus clusters of
galaxies compared with the 2--10 keV total
counts. Datapoints are the same as in Fig.~4.}
\end{figure}

We checked the role of high luminosity absorbed AGNs, the QSO2s,
by fitting the XRB only with unabsorbed AGNs
(at all luminosities) and low luminosity absorbed AGNs.
Following Paper I we introduced an exponential cut off in the 0.5--2 keV 
XLF of absorbed objects, where $L_{s}=2\times10^{44}$ erg s$^{-1}$ is
the intrinsic $e$--folding luminosity. 
The ratio between absorbed and unabsorbed AGNs in the low luminosity regime
is assumed to increase from R(0)=4 to R($z_{cut}$)=10 as in model B.
Then we tuned the parameters of the XLF in order to reproduce 
the XRB spectrum (model C). The fit to the XRB spectrum is
as good as in model B, while a major difference is seen in the 
hard counts (Fig.~7). When trying to reproduce the XRB without QSO2s, 
the 2--10 keV counts at relatively bright fluxes 
are severely underestimated (with a significance higher than 
$4.5\sigma$ at $2\times 10^{-13}$ erg cm$^{-2}$ s$^{-1}$),
since in this case the hard XRB is mainly due to low luminosity objects 
showing up at faint fluxes.

\begin{figure}
\label{hz}
\epsfig{file=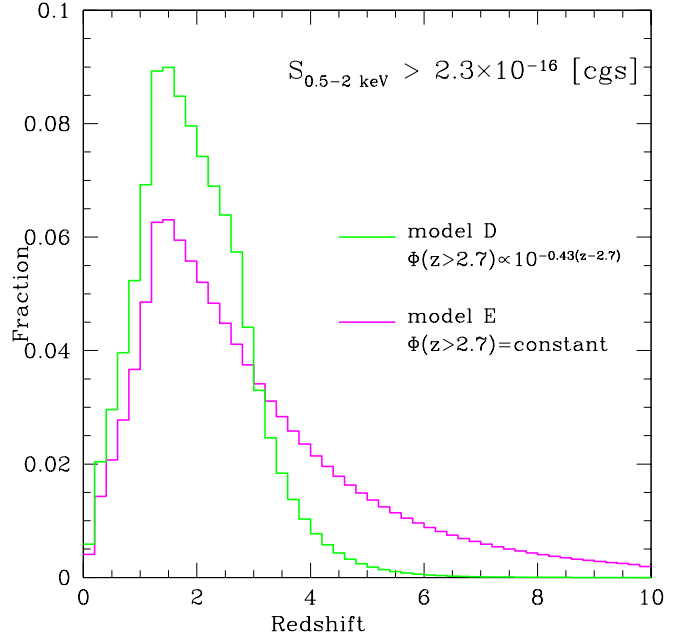, width=9.cm, height=9.cm, angle=0}
\caption{Redshift distribution for the Mushotzky et al. (\cite{musho}) 
0.5--2 keV sample as expected from model D and model E. 
The same distribution is also expected for the 
Giacconi et al. (\cite{giacconi}) 0.5--2 keV sample.}
\end{figure}

\subsection{High--redshift AGNs in X--ray surveys}

Deep surveys should allow to determine the behaviour of the
AGN XLF at high redshift. From optical and radio surveys a decline above
$z\sim3$ is observed, while from X--ray surveys a constant density 
above $z\sim2$ is not ruled out (see Fig.~11 in Mi00a). 
In the previous calculations we assumed that the density of AGNs is
constant in the range from $z=z_{cut}$ to $z=4.6$ and then it suddenly 
drops to zero.
Now, starting from model B, we consider the following scenarios. 
In the first case (model D) we assume that the AGNs density 
above $z=2.7$ decreases by a factor 2.7 per unit redshift as found
for optical QSOs by Schmidt et al. (\cite{schmidt}; 
see also Fan et al. \cite{fan}). 
In the second case (model E)
we assume that the AGNs density is constant up to z=10. 
Model D provides a better description of the
overall X--ray constraints (hard XRB spectrum, soft and hard counts,
soft and hard redshift distributions) than
model E. In particular the Chandra data are better accounted for in model D,
since high redshift (low fluxes) AGNs are removed from the model.
The real behaviour of the AGN density at high redshift will be proven
when complete optical identification of deep surveys will be available.
We calculated the redshift distribution predicted by model D and E at 
$S=2.3\times10^{-16}$ erg cm$^{-2}$ s$^{-1}$, the limiting flux of the
0.5--2 keV sample in Mushotzky et al. (\cite{musho}), which is also very close to the limiting flux of the 0.5--2 keV sample in 
Giacconi et al. (\cite{giacconi}).
The results are shown in Fig.~8. The percentage of AGNs above $z=5$
expected from model D and E are significantly different: 
0.7\% and 15\%, respectively. The total AGN density at
$z>5$ is 16 deg$^{-2}$ for model D and 500 deg$^{-2}$ for model E. 
The Haiman \& Loeb (\cite{haiman}) model
predicts about 500 deg$^{-2}$ AGNs at $z>5$ for the same band and
limiting flux.

\section{Discussion}

In this work some changes have been made in the model assumptions with
respect to Paper I. 
The weakening of the soft excess component slightly hardens the XRB spectrum
and increases the 2--10 keV counts by less than 10\%. In general, our results 
are not significantly affected by the precise parametrization of the soft 
excess. The Compton scattering correction in the spectra of absorbed AGNs with 
log$N_{\rm H}$=24.5 reduces by 10--15\% the intensity of the XRB at 30 keV.
The corrections for the instrumental effects (see the Appendix) are 
relevant on the expected ratio between absorbed and unabsorbed AGNs in 
different surveys, but do not affect the expected total surface density 
of AGNs. The major change, instead, concerns high luminosity obscured AGNs, 
the QSO2s, which are now assumed to constitute a large population of sources.

As already recognized by Comastri (\cite{comastri2}) and as shown in Fig.~7, 
QSO2s are fundamental in reproducing the ASCA counts.
In model C we arbitrarily assume $L_s=2\times10^{44}$ erg s$^{-1}$ as
the intrinsic e--folding luminosity in the 0.5--2 keV XLF of absorbed AGNs.
Objects with this luminosity could still be considered in the Seyfert 
luminosity domain. If we very conservatively assume $L_s=10^{45}$ 
erg s$^{-1}$, corresponding to a bolometric luminosity higher than 
$10^{46}$ erg s$^{-1}$ (Elvis et al. \cite{elvis}), the 2--10 keV counts at 
$2\times 10^{-13}$ erg cm$^{-2}$ s$^{-1}$ are still underestimated 
with a significance higher than $3\sigma$. As shown in Fig.~5 and 6, the
existence and the assumed abundance of QSO2s are consistent with the 
AGN redshift distributions observed in ROSAT and ASCA surveys.

The nature and the number of obscured QSOs is still debated. 
In Paper I, under the 
assumption that the absorbed X--ray photons are re--emitted as infrared 
photons, we estimated that, if every ultraluminous infrared galaxy is
powered by an AGN and is therefore considered as a QSO2, the ratio QSO2s/QSOs
in the local Universe is at most 2. 
This upper limit could be relaxed if a fraction of the 
X--ray absorbed objects have normal blue and broad lined spectrum
in the optical (Reeves \& Turner \cite{reeves}; Ak00; 
Fiore et al. \cite{fiore}).
Radio Loud QSOs were already found to be X--ray absorbed, and the 
X--ray weakness of Broad Absorption Lines QSOs has also been interpreted
as due to X--ray absorption (Brandt et al. \cite{brandt}). 
Furthermore, a large population of X--ray absorbed, broad lined QSOs with
a red continuum seems to emerge from the analysis of grism based surveys 
(Risaliti et al. \cite{risaliti2}) which would join the population of radio 
selected red QSOs discovered by Webster et al. (\cite{webster}). 
However, given the uncertainties affecting the X--ray spectra of these sources,
XMM observations are necessary to confirm the existence of 
a population of broad lined, X--ray absorbed QSOs. 

On the other hand, examples of QSOs obscured both in the optical band 
and in the X rays could be represented by a few ROSAT sources in the Lockman 
Hole with very red colours ($R-K\ga5$, Hasinger et al. \cite{hasinger2}). 
This might 
suggest that the central AGN is heavily obscured and we are 
observing the colours of the host galaxy. As shown by 
Hasinger (\cite{hasinger3}), one 
of these sources, which has been observed from radio to X rays, shows a 
spectral energy distribution remarkably 
similar to that of the luminous absorbed AGN NGC6240 
(Vignati et al. \cite{vignati}).

Overall the number of QSO2s is not yet established, for simplicity we
fixed the local ratio between absorbed and unabsorbed QSOs to 4 as in the case
of lower luminosity AGNs.

As it is apparent from Tab.~2, a model where the evolution of absorbed 
AGNs is faster than that of unabsorbed ones provides a better representation
of the main X--ray observational constraints.
By summing the $\chi^2$ of the 0.5--2 keV XLF, 0.5--2 keV 
logN--logS and 2--10 keV logN--logS, the improvement of model B with 
respect to model A is significant at $>99.99\%$ 
confidence level ($\Delta\chi^2=43.4$). 
Now we discuss some uncertainties which could affect the significance of 
this result. A first problem could be represented by the XRB normalization.
Since there are variations of about 40\% in the 2--10 keV 
XRB intensity measurements (see e.g. Vecchi et al. \cite{vecchi}), 
fitting different
XRB intensities would lead to different predictions for the 
number counts in the soft and hard band. We fitted the XRB intensity 
measured by ASCA, which represents a median level between the highest 
measurement performed by BeppoSAX (Vecchi et al. \cite{vecchi}) and the 
lowest one
performed by HEAO--1 A2 (Gruber \cite{gruber}; Gruber et al. \cite{gruber2}). 
We note that the data 
from the A2 HED experiment on board HEAO--1 seem to be lower by $\sim10\%$ 
not only with respect to the ASCA XRB data, but also with respect to
the high energy data measured by the A4 LED experiment (see Fig.~2 
in Gruber et al. \cite{gruber2}). Thus, a miscalibration of $\sim10\%$ of 
the A2 HED data could account for the some of the systematic differences 
of intensity in the XRB measurements.

In fitting the 2--10 keV XRB intensity measured by 
HEAO--1 A2 by changing the free parameters of the models (see Tab.~1) 
the AGN counts predicted in the soft and hard band would both
be reduced by 20--30\% at the Chandra fluxes, while the decrement is negligible
above $\sim10^{-13}$ erg cm$^{-2}$ s$^{-1}$ and 
$\sim5\times10^{-12}$ erg cm$^{-2}$ s$^{-1}$, respectively. Even in this case
the fast evolution model would provide a better agreement 
with the data.  
Also in the case of a fit to the XRB intensity measured by BeppoSAX model B
provides a better description of the data when compared to model A.  

A second problem could arise from the choices made in the 
$\chi^2$ tests performed on the logN--logS. 
When dealing with the 0.5--2 keV logN--logS we have compared the AGN densities 
predicted by the models with those measured by ROSAT. Since for the Chandra 
data the optical identifications are sparse, we compared the predictions
of the model with the total number of sources. 
We note that in our models the cluster contribution at the Chandra fluxes
is negligible.
However, if other classes of sources are contributing to the Chandra 
counts and the AGNs counts are significantly lower than the total ones, 
model B would always provide a better representation of the data.

When considering the 2--10 keV logN--logS we compared the data with the
total density of objects (AGN plus clusters of galaxies) 
predicted by the models, since 
optical identifications are incomplete for most of the adopted datasets.
We note however that our models are in agreement 
with the separate measurements of the AGN and cluster densities determined by 
Piccinotti et al. (\cite{picci}) and Ak00.
We consider it appropriate to compare the total source density measured by the 
different surveys in the 2--10 keV band with the AGN + cluster model 
predictions, since $>95\%$ of the sources in the Ak00
sample are AGNs or clusters. 
We note that without the correction for the ASCA instrumental response
the ratio N(log$N_{\rm H}<22$)/N(log$N_{\rm H}>22$) predicted by model B for the Ak00
sample would have been
12/18 rather than 15/15. As pointed out by Comastri (\cite{comastri2}), 
in previous synthesis models the number
of unabsorbed AGNs predicted at the ASCA fluxes
seemed to be in disagreement with that observed. When considering 
instrumental effects much of the discrepancy can thus be explained.

After the discovery of hard X--ray spectra in a small sample of nearby 
elliptical galaxies (Allen, Di Matteo \& Fabian \cite{allen}) 
it was proposed by Di Matteo \& Allen (\cite{dimatteo}) that a 
substantial fraction of the XRB intensity could arise from advection dominated
flows in elliptical galaxies. However, in this case elliptical galaxies with
normal optical spectra should represent a significant fraction of the 2--10 keV
ASCA counts, at variance with the findings of Ak00.

Our models are found to be in good agreement with the AGN redshift 
distributions in the soft and hard bands, as checked with the 
Kolmogorov--Smirnov tests. Also, they are in agreement with the 
relative fraction of absorbed AGNs found in flux limited samples. 
A possible discrepancy could be recognized 
in the ratio between AGNs with log$N_{\rm H}<22$
and log$N_{\rm H}>22$ observed in the Ak00 sample and that predicted 
by model B. However, this is only a $1\sigma$ effect. Also, 
the column density estimated by Ak00 from the hardness ratio analysis 
assuming a simple absorbed power law model could have been underestimated 
if a soft component is commonly present in the spectra of absorbed AGNs, as
assumed in our models. This is likely to be the case, as shown by the X--ray 
color--color diagrams for ASCA and BeppoSAX sources (Della Ceca et al. 
\cite{rdc}; Fiore et al. \cite{fiore}).

Our models slightly overpredict the Chandra source counts in the soft
band, and probably also in the hard band (referring to the data of Giacconi 
et al. \cite{giacconi}). Introducing a cut off at $z=2.7$ in the AGN density as
observed in the optical surveys would provide a better description 
of the Chandra data, since high redshift AGNs which would show up at faint
fluxes would be removed. 

The improvement introduced by model B with respect to model A resides 
essentially
in the higher fraction of hard sources and in the lower value of $z_{cut}$.
In order to avoid introducing other free parameters we assumed that the
value of $z_{cut}$ is the same for absorbed and unabsorbed AGNs. However, 
we note that a still better fit to the X--ray constraints could be obtained if 
a lower $z_{cut}$ for absorbed AGNs were used. Indeed, this would imply 
that a higher fraction of the hard XRB is produced at small redshift, with 
an increase of the hard counts at the ASCA fluxes and a decrease at the 
Chandra fluxes. Also, the XRB bump at $\sim 30$ keV would be better
reproduced since the AGN spectral peak at 30--40 keV would be on the average
less redshifted. Another improvement is expected in the redshift distributions.

We note that the $N_{\rm H}$ distribution adopted in our models, derived 
by Risaliti et al. (\cite{risaliti}) for a local sample of AGNs, is very 
similar\footnote{Basically the two distributions differ only about the 
number
of AGNs with log$N_{\rm H}>25$ (see Fig.~1 of Paper I), which however do not
contribute significantly to the XRB intensity.} to that used by Comastri
et al. (\cite{comastri}) {\it to fit} the XRB spectrum,
which is mostly produced by objects at higher redshift. It thus seems that
it would be difficult to obtain good XRB fits with $N_{\rm H}$ distributions 
which are very different from the adopted one. Furthermore, this argument
suggests that a fast evolution of absorbed sources would be hardly superseded
in improving the description of X--ray data by simply changing the 
$N_{\rm H}$ distribution with redshift.

The significance of a faster evolution of absorbed AGNs might be 
reduced by systematic uncertainties in the adopted data and further 
checks with the data from on going deep surveys are needed.
If the higher fraction of absorbed AGNs at high redshift is proven to be 
real, one could try to link this finding to the star formation history in 
the Universe. A higher merging rate at high redshift could trigger the star 
formation and the gas flow towards the galaxy centers. 
As suggested by Fabian et al. (\cite{fabian}), circumnuclear starbursts
could obscure most of the nuclear radiation. 
In this scheme, the higher fraction  
of obscured sources at high redshift could be explained. 
This can also be related with the finding that non--axisymmetric 
morphologies increase the obscuration in AGNs 
(Maiolino, Risaliti \& Salvati \cite{maio2}), 
and that distorted, non--axisymmetric morphologies are more common among 
galaxies at high redshift.

\section{Conclusions}

We have presented different synthesis models for the XRB, checking the 
model predictions with all the available X--ray constraints posed by
the source counts, XLF, redshift distributions and absorption
distributions. Models assuming a population of high luminosity obscured 
AGNs, the QSO2s, are found to be consistent with all the available data. 
Furthermore, the existence of QSO2s is found to be necessary in reproducing 
the 2--10 keV source counts at relatively bright fluxes 
($\sim 10^{-13}$ erg cm$^{-2}$ s$^{-1}$). 
We found that a model (model B) where the evolution of absorbed AGNs is 
faster than that of unabsorbed ones provides a better description of the
data with respect to a standard model (model A) where absorbed and unabsorbed 
AGNs evolve with the same rate. 
Our models are also in agreement with the first results from
Chandra deep surveys. Chandra sources, together with those detected in XMM 
deep surveys, will provide additional informations on the QSO2 population 
and will verify if the faster evolution of absorbed AGNs is real. 
Furthermore, the behaviour of the AGN space density
at high redshift will be determined when optical identification programs
will be completed.

\begin{acknowledgements}

We are grateful to T. Miyaji for providing data on the XRB spectrum 
and AGN XLF, and to P. Tozzi for providing data on the Chandra counts. 
RG thanks A. Comastri and G. Risaliti for useful discussions.
We thank the referee for prompt and constructive comments.
This work was partly supported by the Italian Space Agency
(ASI) under grant ARS--98--116/22 and by the Italian Ministry for
University and Research (MURST) under grant Cofin98--02--32. 

\end{acknowledgements}

\appendix
\section{Corrections for instrumental effects}

To allow a consistent comparison between the logN--logS
and redshift distributions predicted by the models and the literature data,
we have taken into account the bias against absorbed sources introduced by 
the effective area of the X--ray instruments.
We describe the procedure adopted when comparing the model predictions with
the 0.5--2 keV AGN logN--logS, which is mainly derived from ROSAT PSPC data.
This procedure is analogous to that adopted when considering data from other 
X--ray instruments. 
For each absorption class considered in the models we have produced with 
the XSPEC package a spectrum identical to that considered in the model.
The obtained spectra have been folded with the ROSAT PSPC spectral
response. We adopted the calibration
file {\tt pspcb\_gain2\_256.rsp} retrieved from the archive in the
ftp://legacy.gsfc.nasa.gov/ website. The file contains the spectral response 
of the PSPC detector multiplied by the X--ray telescope effective
area.
We therefore obtained a count rate to flux conversion factor CF$_i$ for  
each of the AGN spectra assumed in our model. The spectra have also been
redshifted and the conversion factors calculated as a function of redshift.
When deriving the AGN 0.5--2 keV XLF and counts Mi00a
assumed a power law with $\Gamma=2$ (absorbed by a Galactic column density) 
to convert from the 0.5--2 keV PSPC count rate to the 0.5--2 keV flux, 
corresponding to a conversion factor 
CF$_*=0.836$ cts s$^{-1}$/ ($10^{-11}$ erg cm$^{-2}$ s$^{-1}$) 
(see Tab.~2 in Hasinger et al. \cite{hasinger}).
The conversion factor CF$_*$ was used for all the sources, irrespectively
of their real spectrum. However, sources with a spectrum harder than 
$\Gamma=2$ would produce less photons on the detector at a given flux, 
therefore a conversion factor lower than CF$_*$ should be used. 
Analogously, for those sources softer than $\Gamma=2$ a conversion factor 
greater that CF$_*$ should be used. 
Then, a source with a measured flux S would have a real flux
S$\times$(CF$_*$/CF), where CF is the conversion factor which would 
be appropriate for the source spectrum.
Therefore, in our calculations of the logN--logS, at a given flux S we summed 
the contribution of each absorption class of the model calculated at 
S$\times$(CF$_*$/CF$_i$).

The same procedure has been adopted for the 2--10 keV logN--logS and
redshift distribution at S=$10^{-13}$ erg cm$^{-2}$ s$^{-1}$, which have 
been corrected for the ASCA SIS0 plus X--ray mirror effective area 
(calibration file {\tt s0c1g0234p40e1\_512\_1av0\_8i.rsp}). 
For the 2--10 keV band we adopted 
CF$_*=1.089$ cts s$^{-1}$/ ($10^{-10}$ erg cm$^{-2}$ s$^{-1}$),
corresponding to $\Gamma=1.7$. Indeed, a conversion spectrum with 
$\Gamma=1.6-1.7$ is commonly adopted in the hard X--ray surveys 
(e.g. Ak00; Cagnoni et al. \cite{cagnoni}).

For the redshift distribution of the 0.5--2 keV AGNs calculated  
at S=$1.2\times10^{-15}$ erg cm$^{-2}$ s$^{-1}$
we have considered the corrections for the HRI spectral response and X--ray 
mirror effective area contained in {\tt hri\_90dec01.rsp}. 
We adopted CF$_*=0.586$ cts s$^{-1}$/ ($10^{-11}$ erg cm$^{-2}$ s$^{-1}$),
which is the conversion factor from the 0.1--2.4 keV HRI cont rate to the 
0.5--2 keV flux appropriate for a spectral power law with $\Gamma=2$.


\begin{thebibliography}{}

\bibitem[2000]{ak00} Akiyama M., Ohta K., Yamada T., et al., 2000, ApJ 532, 
700 (Ak00)

\bibitem[2000]{allen} Allen S.W., Di Matteo T., Fabian A.C., 2000, MNRAS 311, 
493

\bibitem[2000]{barger} Barger A.J., Cowie L.L., Mushotzky R.F., Richards E.A.,
2000, AJ submitted [astro--ph/0007175]

\bibitem[1996]{bower} Bower R.G., Hasinger G., Castander F.J., et al., 1996, 
MNRAS 281, 59

\bibitem[2000]{brandt} Brandt W.N., Laor A., Wills B.J., 2000, ApJ 528, 637

\bibitem[1998]{cagnoni} Cagnoni I., Della Ceca R., Maccacaro T., 1998, 
ApJ 493, 54

\bibitem[1995]{comastri} Comastri A., Setti G., Zamorani G., Hasinger G., 1995,
A\&A 296, 1 

\bibitem[2000]{comastri2} Comastri A., 2000, Astroph. Lett. and Comm., in press
[astro--ph/0003437]

\bibitem[1999]{rdc} Della Ceca R., Castelli G., Braito V., Cagnoni I., 
Maccacaro T., 1999, ApJ 524, 674

\bibitem[2000]{rdc2} Della Ceca R., Braito V., Cagnoni I., Maccacaro T., 2000,
Mem. SAIt, in press [astro--ph/0007430]

\bibitem[1999]{dimatteo} Di Matteo T., Allen S.W., 1999, ApJ 527, L21

\bibitem[1994]{elvis} Elvis M., Wilkes B.J., Mc Dowell C., et al., 1994, 
ApJS 95, 1 

\bibitem[1998]{fabian} Fabian A.C., Barcons X., Almaini O., Iwasawa K., 1998, 
MNRAS 297, L11

\bibitem[2000]{fan} Fan X., Strauss M.A., Schneider D.P., et al., 2000, 
AJ, in press [astro--ph/0008123]

\bibitem[2000]{fiore} Fiore F., Giommi P., Vignali C., et al., 2000, 
MNRAS, submitted

\bibitem[1995]{gendreau} Gendreau K.C., Mushotzky R.F., Fabian A.C., et al., 
1995, PASJ 47, L5

\bibitem[2000]{george} George I.M., Turner T.J., Yaqoob T., et al., 2000, 
ApJ 531, 52

\bibitem[2000]{giacconi} Giacconi R., Rosati P., Tozzi P., et al., 2000, 
ApJ, submitted [astro--ph/0007240]

\bibitem[1999]{gilli} Gilli R., Risaliti G., Salvati M., 1999, A\&A 347, 424

\bibitem[2000]{giommi} Giommi P., Perri M., Fiore F., 2000, A\&A, in press 
[astro--ph/0006333]
 
\bibitem[1992]{gruber} Gruber D.E., 1992,
In: Barcons X., Fabian A.C. (eds.) The Proceedings of: The X--ray
background. Cambridge Univ. Press, Cambridge, p.44

\bibitem[1999]{gruber2} Gruber D.E., Matteson J.L., Peterson L.E., Jung G.V., 
1999, ApJ 520, 124

\bibitem[1999]{haiman} Haiman Z., Loeb A., 1999, ApJ 521, L9

\bibitem[1999]{halpern} Halpern J.P., Turner T.J., George I.M., 1999, 
MNRAS 307, L47

\bibitem[1998]{hasinger} Hasinger G., Burg R., Giacconi R., Schmidt M., 
Tr\"umper J., Zamorani G., 1998, A\&A 329, 482

\bibitem[1999]{hasinger2} Hasinger G., Lehmann I., Giacconi R., et al.,
1999, In: Highlights in X--ray Astronomy in Honor 
of Joachim Tr\"umper's 65th Birthday. MPE Report, MPE, Garching, in press
[astro--ph/9901103]

\bibitem[2000]{hasinger3} Hasinger G., 2000, In: Lemke D., Stickel M., Wilke K.
(eds.) The Proceedings of: ISO Surveys of a Dusty Universe. Springer, in press 
[astro--ph/0001360]

\bibitem[1998]{kim} Kim D.--C., Sanders D.B., 1998, ApJS 119, 41 

\bibitem[1994]{madau} Madau P., Ghisellini G., Fabian A.C., 1994, 
MNRAS 270, L17

\bibitem[1995]{maio} Maiolino R., Rieke G.H., 1995, ApJ 454, 95

\bibitem[1999]{maio2} Maiolino R., Risaliti G., Salvati M., 1999, A\&A 341, L35

\bibitem[2000]{maio3} Maiolino R., Marconi A., Salvati M., et al., 2000, 
A\&A, in press [astro--ph/0010009]

\bibitem[2000]{mason} Mason K.O., Carrera F.J., Hasinger G., et al., 2000, 
MNRAS 311, 456

\bibitem[1999]{matt} Matt G., Pompilio F., La Franca F., 1999, New Astron. 4, 
191

\bibitem[2000a]{miyaji} Miyaji T., Hasinger G., Schmidt M., 2000a, A\&A 353, 
25 (Mi00a)

\bibitem[2000b]{miyaji2} Miyaji T., Hasinger G., Schmidt M., 2000b, 
A\&A, submitted

\bibitem[2000]{musho} Mushotzky R.F., Cowie L.L., Barger A.J., Arnaud K.A., 
2000, Nat. 404, 459

\bibitem[1998]{ogasaka} Ogasaka Y., Kii T., Ueda Y., et al., 1998, Astron. 
Nacht. 319, 47

\bibitem[1996]{page} Page M.J., Carrera F.J., Hasinger G., 1996, 
MNRAS 281, 579

\bibitem[1982]{picci} Piccinotti G., Mushotzky R.F., Boldt E.A., et al., 
1982, ApJ 253, 485

\bibitem[2000]{pompilio} Pompilio F., La Franca F., Matt G., 2000, A\&A 353, 
440

\bibitem[2000]{reeves} Reeves J.N., Turner M.J.L, 2000, MNRAS 316, 234 

\bibitem[1997]{reynolds} Reynolds C.S., 1997, MNRAS 286, 513

\bibitem[1999]{risaliti} Risaliti G., Maiolino R., Salvati M., 1999, ApJ 522, 
157

\bibitem[2000]{risaliti2} Risaliti G., Marconi A., Maiolino R., Salvati M., 
Severgnini P., 2000, A\&A, submitted   

\bibitem[1997]{schartel} Schartel N., Schmidt M., Fink H.H., Hasinger G., 
Tr\"{u}mper J., 1997, A\&A 320, 696

\bibitem[1995]{schmidt} Schmidt M., Schneider D.P., Gunn J.E., 1995, AJ 110, 68

\bibitem[1998]{schmidt2} Schmidt M., Hasinger G., Gunn J., et al.,
1998, A\&A 329, 495

\bibitem[1989]{setti} Setti G., Woltjer L., 1989, A\&A 224, L21

\bibitem[1999]{shaver} Shaver P.A., Hook I.M., Jackson C.A., et al., 1999, 
In: Carilli C., Radford S., Menten K., Langston G., (eds.) Highly Redshifted 
Radio Lines. ASP Conf. Series Vol. 156, p.163 [astro--ph/9801211]

\bibitem[1999]{ueda} Ueda Y., Takahashi T., Inoue H., et al., 1999, ApJ 518, 
656

\bibitem[1999]{vecchi} Vecchi A., Molendi S., Guainazzi M., Fiore F., 
Parmar A.N., 1999, A\&A 349, L73

\bibitem[1999]{vignati} Vignati P., Molendi S., Matt G., et al.,  1999, A\&A 
349, L57

\bibitem[1999]{veilleux} Veilleux S., Kim D.--C., Sanders D.B., 1999, ApJ 522,
113

\bibitem[1995]{webster} Webster R.L., Francis P.J., Peterson B.A., et al., 
1995, Nat. 375, 469

\bibitem[1999]{gz} Zamorani G., Mignoli M., Hasinger G., et al., 1999, A\&A 
346, 731

\end{thebibliography}
\end{document}